\documentclass[letterpaper,english,aps,prl,preprintnumbers,amsmath,amssymb,twocolumn,superscriptaddress,showpacs,floatfix]{revtex4}
\usepackage[T1]{fontenc}
\usepackage[latin9]{inputenc}
\usepackage{mathrsfs}
\usepackage{amsmath}
\usepackage{amssymb}
\usepackage{graphicx}
\usepackage{esint}
\usepackage{color}
\usepackage{ulem}
\makeatletter
\@ifundefined{textcolor}{}
{%
 \definecolor{BLACK}{gray}{0}
 \definecolor{WHITE}{gray}{1}
 \definecolor{RED}{rgb}{1,0,0}
 \definecolor{GREEN}{rgb}{0,1,0}
 \definecolor{BLUE}{rgb}{0,0,1}
 \definecolor{CYAN}{cmyk}{1,0,0,0}
 \definecolor{MAGENTA}{cmyk}{0,1,0,0}
 \definecolor{YELLOW}{cmyk}{0,0,1,0}
}
\makeatother

\usepackage[colorlinks,
            linkcolor=blue,
            anchorcolor=blue,
            citecolor=blue
            ]{hyperref}

\usepackage{babel}

\begin{document}

\title{Effect of medium recoil and $p_T$ broadening on single inclusive jet suppression in high-energy heavy-ion collisions
{in the high-twist approach}}

\author{Xin-Nian Wang}
\affiliation{Key Laboratory of Quark and Lepton Physics (MOE) and Institute of Particle Physics, Central China Normal University, Wuhan 430079, China}
\affiliation{Nuclear Science Division Mailstop 70R0319,  Lawrence Berkeley National Laboratory, Berkeley, California 94740, USA}

\author{Shu-Yi Wei}
\affiliation{Key Laboratory of Quark and Lepton Physics (MOE) and Institute of Particle Physics, Central China Normal University, Wuhan 430079, China}

\author{Han-Zhong Zhang}
\affiliation{Key Laboratory of Quark and Lepton Physics (MOE) and Institute of Particle Physics, Central China Normal University, Wuhan 430079, China}

\begin{abstract}
Jet energy loss and single inclusive jet suppression in high-energy heavy-ion collisions are studied within a pQCD parton 
model that includes both elastic and radiative interactions between jet shower and medium partons as they 
propagate through the quark-gluon plasma. The collisional energy loss of jets with a given cone-size is found to be relatively small 
comparing with the radiative energy loss. However the effect of transverse momentum broadening due to elastic scattering is significant 
in the calculation of radiative energy loss within the high-twist formalism. The nuclear modification factors for single inclusive jets with
different cone-sizes are calculated and compared to experimental data as measured by ALICE and ATLAS experiments in Pb+Pb collisions
at $\sqrt{s_{\rm NN}}=2.76$ TeV. Results on jet suppression in Pb+Pb collisions at $\sqrt{s_{\rm NN}} = 5.02$ TeV are also presented.
\end{abstract}
\maketitle

\section{Introduction}
In high-energy heavy-ion collisions, jet quenching phenomena \cite{Gyulassy:1990ye,Wang:1991xy} can be used as a powerful tool to 
study properties of the strongly interacting quark-gluon plasma (QGP). These phenomena are caused by parton energy loss and transverse
momentum $p_T$ broadening due to interaction between jet shower partons and thermal medium partons \cite{Gyulassy:1993hr, Baier:1994bd, Baier:1996sk, Zakharov:1996fv, Gyulassy:2000fs, Gyulassy:2000er, Wiedemann:2000za, Wiedemann:2000tf, Guo:2000nz, Wang:2001ifa,Wang:2009qb, Qin:2015srf}. 
%
%
{
The radiative energy loss can be related to the jet transport coefficient $\hat q$ that characterizes the properties of the dense medium \cite{Baier:1994bd, Baier:1996sk}.
At the leading order in pQCD, $\hat q$ is directly the averaged transverse momentum broadening squared per unit length. The evaluation of $\hat q$ is model dependent and could involve non-perturbative physics. Recently, it has been studied at the next-to-leading order (NLO) in pQCD \cite{Wu:2014nca, Kang:2016ron}. The elastic energy loss is however not directly related to $\hat q$ but another jet transport coefficient $\hat e$ \cite{Majumder:2008zg}.
}
Recent phenomenological studies have extracted the value of the jet transport coefficient from experimental data on suppression of single inclusive hadron spectra at large transverse momentum \cite{Burke:2013yra}.  Studies of reconstructed jets and their medium modification should provide further constraints on the properties the dense medium \cite{Vitev:2009rd,Wang:2013cia,Qin:2010mn,CasalderreySolana:2010eh,Lokhtin:2011qq,Young:2011qx,He:2011pd,Renk:2012cx,CasalderreySolana:2012ef,Dai:2012am,Qin:2012gp,Chien:2015hda,Chang:2016gjp} in heavy-ion collisions at the RHIC, LHC and future high-energy collider energies \cite{Chang:2015hqa}.

In model descriptions of the jet quenching phenomena in both high $p_T$ hadron and jet spectra, energy loss due to elastic and inelastic processes
should be both considered \cite{Chang:2016gjp,Wicks:2005gt,Qin:2007rn,Schenke:2009ik}. Though elastic energy loss is significantly smaller than radiative energy loss for light partons, it plays an important role  in providing a consistent description of suppression of both light and heavy quark hadrons in high-energy heavy-ion collisions \cite{Cao:2013ita,Cao:2015hia, Djordjevic:2014hka,Djordjevic:2013xoa}.  

{
In the study of net energy loss for reconstructed jets with given jet cone-size, we should consider the scenario when recoil partons from the medium in the elastic process fall inside the jet cone. In this case, the total energy within the jet-cone remains the same and therefore no net energy loss for the jet. The net energy loss for jets with finite cone-size considering recoil partons is therefore different with the elastic energy loss of a single parton.
}
Such recoil effect from jet medium interaction has been found important in Monte Carlo simulations of jet propagation and modification \cite{Luo:2014aia,He:2015pra,Casalderrey-Solana:2015vaa}. The diffusion of jet shower partons in the transverse direction should also influence the net jet energy loss and the jet transverse profile \cite{Chang:2016gjp,He:2015pra,Casalderrey-Solana:2015vaa}.

In this paper, we will study the effects of recoil partons and transverse momentum broadening on the net jet parton energy loss and single inclusive jet suppression within the NLO perturbative QCD parton model in high-energy heavy-ion collisions. We will consider different scenarios of recoil thermal parton propagation after the elastic scatterings with the jet shower partons. Within the high-twist approach to radiative parton energy loss, the momentum of radiated gluons are most likely to be collinear with that of the initial parton. For jets with a certain cone-size,  small angle radiations that fall inside the jet cone do not contribute to the jet energy loss. The diffusion of jet shower partons and radiated gluons due to transverse momentum broadening should have significant effects on net jet energy loss within the jet cone. We will include transverse momentum broadening of both jet shower and radiated partons and study their effects on net jet energy loss and suppression of single inclusive jet spectra.

\section{Collisional energy loss and recoil partons}

The energy loss rate of a jet with jet-cone size $R$ due to elastic scattering can be calculated from the elastic scattering rate \cite{He:2015pra},
\begin{eqnarray}
\frac{dE_a^{\rm el}}{dx} &\equiv & \sum_{b,c,d} \int d\theta_2 d\theta_3 d\phi_{3} dE_3 
f_b (E_2,T) \frac{|\mathcal{M}_{ab\to cd}|^2}{16E_1(2\pi)^4}\nonumber \\
& & \times\frac{E_2E_3\sin \theta_2 \sin \theta_3  \delta E_{ab\to cd}}{[E_1 (1-\cos\theta_2) - E_3 (1-\cos\theta_{3})]} 
\end{eqnarray}
where
\begin{equation}
{{E}_{2}}=\frac{{{E}_{1}}{{E}_{3}}(1-\cos {{\theta }_{3}})}{{{E}_{1}}(1-\cos {{\theta }_{12}})-{{E}_{3}}(1-\cos {{\theta }_{23}})},
\end{equation}
$\phi_i$ is the azimuthal angle,  $\theta_i$ is the polar angle of a parton's momentum $p_i$,  $\phi_{ij}$ and  $\theta_{ij}$  are  the azimuth and polar angles between two partons' momenta $p_i$ and $p_j$, respectively. The thermal parton distribution function in the medium, $f_b$ is given by the Fermi-Dirac distribution for a quark, or the Bose-Einstein distribution for a gluon. The matrix elements $|\mathcal{M}_{ab\to cd}|^2$ for two-parton scatterings \cite{Eichten:1984eu} are regularized with a Debye mass $\mu_D^2=(1+n_f/6)g^2T^2$.

To calculate the net energy loss of a jet with a finite jet-cone size, one should take into account of the recoil thermal partons before and after the elastic scattering \cite{He:2015pra}. The thermal parton can become a part of the jet if it falls inside the jet-cone after the elastic scattering. One should also subtract the initial thermal parton's energy as part of the background if it was within the jet-cone before the elastic scattering. The net energy loss within the jet-cone due to elastic scattering is therefore,
\begin{equation}
\delta E^{\rm el}_{\rm w/recoil} = E_a + E_b \theta_b - E_c \theta_c - E_d \theta_d,
\end{equation}
where $\theta_i$ is a $\theta$-function in the relative angle between partons and the jet: it equals to 1 if the parton falls inside the jet cone, 0 otherwise. In this scenario, we neglect further interaction of the recoil thermal partons.

In an alternative scenario, we assume the recoil parton is immediately thermalized and become part of the medium. The contribution of the thermalized recoil parton to the net jet energy within the jet-cone is neglected. The net collisional jet energy loss per scattering in this scenario is,
\begin{equation}
\delta E^{\rm el}_{\rm n/recoil} = E_a - {\rm max}(E_c,E_d),
\end{equation}
which is just the elastic energy loss of a singe parton.

\begin{figure}[htb]
\includegraphics[width=0.45\textwidth]{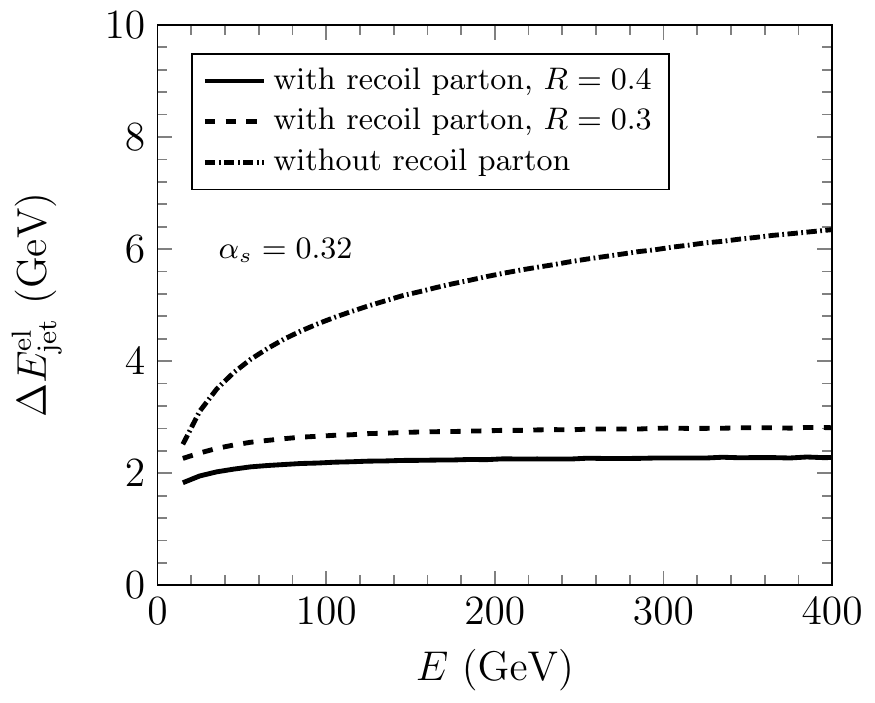}
\caption{Collisional energy losses for a quark jet that is produced at $x=0$, $y=0$, $\phi=0$ and propagates though the medium in the $\sqrt{s_{\rm NN}} = 2.76$ TeV Pb+Pb collisions in the $0-10\%$ centrality.}
\label{fig:radiative.and.collisional}
\end{figure}

To illustrate the effect of recoil partons in the collisional energy loss, we calculate the total elastic energy loss for a quark jet that is initially produced at $x=y=z=0$ and propagates in the transverse direction along the direction $\phi=0$ through the QGP produced in Pb+Pb collisions at $\sqrt{s_{\rm NN}}=2.76$ TeV with $0-10\%$ centrality ($b=3.2$ fm). The total collisional energy loss for a jet propagating though the medium is the integral of the collisional energy loss rate over the propagating path, 
$\Delta E^{\rm el}_{\rm jet} = \int_{\rm path} dx dE^{\rm el}/dx$.
The temperature profile for determining the local thermal parton distributions along the propagation path is given by a hydrodynamic
evolution \cite{Hirano:2005xf,Hirano:2007ei} with the initial temperature $T_0 = 468$ MeV at $x=0$, $y=0$ and $\tau_0=0.6$ fm.
In the leading order approximation of a reconstructed jet, we assume the jet-cone size $R$ is defined as the opening angle with respect to the initial parton before the elastic scattering.  Shown in Fig.~\ref{fig:radiative.and.collisional} are the total net jet energy loss due to elastic scattering with (solid and dashed lines) and without recoil thermal partons (dot-dashed lines) as a function of the initial jet transverse momentum $p_T$ for a jet cone-size $R=0.4$ and $R=0.3$. It is clear that inclusion of recoil partons inside the jet cone reduces the net elastic jet energy loss significantly, especially for a large jet cone size.

One should note that the recoil thermal partons from the elastic jet-medium scattering will go through further scattering in the medium. Though the net jet energy loss would not change much due to these further interaction which is dominated by small angle scattering, the energy distribution at large angles in the jet profile is very sensitive to the further propagation of recoil thermal partons. This effect has been studied in the Linear Boltzmann Transport (LBT) model \cite{He:2015pra,Luo:2014aia,Luo:2016ruj}, the Boltzmann Approach to Multi-Parton Scatterings (BAMPS) model \cite{Senzel:2013dta, Senzel:2016qau} and the hybrid model \cite{Casalderrey-Solana:2015vaa} in which energy loss to the medium is propagated via hydrodynamic evolution \cite{Tachibana:2015qxa}. We will neglect this effect due to recoil parton propagation in this study since we are only concerned with the net jet energy loss in the suppression of single inclusive jet cross section in high-energy heavy-ion collisions.

\section{Radiative energy loss with $p_T$ broadening}

Jet shower partons propagating in the QGP medium will also suffer radiative energy loss due to gluon bremsstrahlung in addition to the elastic energy loss.  It is the dominant mechanism for parton energy loss in the study of suppression of single hadron spectra \cite{Burke:2013yra,Wang:2004dn,Vitev:2002pf,Wang:2003mm,Eskola:2004cr,Renk:2006nd,Zhang:2007ja} and jet spectra \cite{Vitev:2009rd,Wang:2013cia,Qin:2010mn,CasalderreySolana:2010eh,Lokhtin:2011qq,Young:2011qx,He:2011pd,Renk:2012cx,CasalderreySolana:2012ef,Dai:2012am,Qin:2012gp,Chien:2015hda,Chang:2016gjp} in high-energy heavy-ion collisions. 

In most theoretical studies, radiative parton energy loss is dominated by soft and collinear gluon radiation \cite{Gyulassy:1993hr, Baier:1994bd, Baier:1996sk, Zakharov:1996fv, Gyulassy:2000fs, Gyulassy:2000er, Wiedemann:2000za, Wiedemann:2000tf, Guo:2000nz, Wang:2001ifa}. Further scatterings of the radiative gluon are neglected since they don't cause additional energy loss of the leading parton. For the net energy loss of reconstructed jets with finite jet-cone size, interaction of the radiative gluons with the medium during their propagation can transport energy into or out of the jet cone and therefore should be considered. We will consider the transverse momentum broadening of radiative gluons and its effect on the final net jet energy loss in this paper.

We will work within the high-twist model in which parton energy loss can be approximately expressed as \cite{Guo:2000nz, Wang:2001ifa,Wang:2009qb},
\begin{eqnarray}
 \frac{\Delta E_{\rm a}^{\rm rad} }{E}& \approx &  
\frac{2}{\pi} C_A   \alpha_s \int d\tau \int_0^{0.5} dz \int \frac{dl_T^2}{l_T^4}  \nonumber \\
&\times& \hat q_a z P_{ga} (z)  \sin^2 \left( \frac{l_T^2 \tau}{4z(1-z)E} \right),
\label{eq:rad.lp}
\end{eqnarray}
for parton $a$ with initial energy $E$, where  $\alpha_s$ is the strong coupling constant, $z$ is the energy fraction of the radiated gluon, 
$l_T$ is its transverse momentum, and $\int d\tau$ is the integral along the propagation path.
$P_{ga}(z)$ is the splitting function without the color factor. For a quark jet 
$P_{gq}(z) = [1 + (1-z)^2]/z$.
{We take the small $z$ approximation by limiting the upper limit of the integration over $z$ to be $1/2$. In this case, after the splitting, the flavor of the leading parton has not been changed. This gives the clear physical picture of energy loss.
This is also consistent with Eq.~(\ref{eq:rad.lp}). We can see from Eq.~(\ref{eq:rad.lp}), the dominate contribution to the parton energy loss comes from the small-$z$ region.} 
The jet transport parameter $\hat q_a$ or the average transverse momentum broadening squared per unit length
will be given by leading order perturbative elastic scattering between parton $a$ and the medium \cite{He:2015pra},
\begin{align}
\hat q_a= C_a\frac{42\zeta (3)}{\pi} \alpha_s^2 T^3 
\ln \left( \frac{s^*}{4\mu_D^2} \right),
\end{align}
where $C_a=C_F=\frac{4}{3}$, $s^* = 5.8ET$ for a quark and $C_a=C_A=3$, $s^* = 5.6ET$ for a gluon, 
$\mu_D$ is the Debye mass,  $\zeta (3) \approx 1.202$ is the Ap\'{e}ry's constant. The local temperature $T$ along the propagation path at time $\tau$ will be given by the hydro profile of
heavy-ion collisions with the given centrality \cite{Hirano:2005xf,Hirano:2007ei}.
For the radiative energy loss of a gluon jet, one should replace the quark splitting function $P_{gq}$ with the gluon one $P_{gg}$. The difference between the numerical results given by these two splitting functions is less than one percent except the color factor which has been absorbed into $\hat q_a$. {The $g\to q\bar q$ splitting can be ignored in the small $z$ approximation.}

To calculate the net jet energy loss within a jet cone, one has to include three modifications based on the above description of parton energy loss. The first modification is to consider jet cone size. For radiated gluons that are collinear with the initial leading parton, they don't contribute to the net jet energy loss if they are still inside the jet cone. Only those radiated gluons that fall outside the jet cone contribute to the net jet energy loss,
\begin{equation}
l_T / zE > \sin (R), \label{cut:collinear.unbro}
\end{equation}
where $R$ is the cone-size.

The second modification involves soft radiated gluons. We assume those soft radiated gluons whose energy is less than the Debye mass
$\mu_D \sim gT$ become thermalized with the medium through further interaction and their contribution to the jet energy within the jet cone is negligible.  Hence, the energy carried by the soft radiated gluons, $zE<\mu_D \sim 1{\rm GeV}$ is considered lost to the medium regardless of their transverse momentum.

The third modification comes from the transverse momentum broadening while a jet traverses the medium. 
Within the high-twist approach to radiative energy loss, collinear approximation is made through a Taylor expansion in the
transverse momentum of exchanged gluons. At twist-4, the transverse momentum enters the final results through an averaged
quantity, the jet transport coefficient $\hat q$. However, the final radiated gluon in the high-twist approach does not carry any
transverse momentum broadening from the interaction with the medium. 
{
In this paper, we explore the broadening effect of the radiated gluon using a Gaussian broadening model.
}

The diffusion of radiated gluons in the transverse direction due to transverse momentum broadening will lead to transport of radiated gluons outside of the jet cone and therefore increase the net jet energy loss. Such transport of radiated gluons has been taken into account in LBT model \cite{Wang:2013cia} and other parton transport Monet Carlo models \cite{Zapp:2008gi,Schenke:2009gb}. It is therefore important to take into account of this effect in pQCD parton models with medium induced parton energy loss. Within the high-twist framework, one can assume the transverse momentum distribution due to $p_T$ broadening takes a Gaussian form. The averaged transverse momentum broadening squared is determined by the path integral of the jet transport coefficient $\hat q$,
\begin{equation}
\langle \Delta l_T^2 \rangle \equiv \int_{\tau_0}^{\infty} d\tau \hat q (\vec r_0 + \vec v \tau).
\end{equation}
With this additional transverse momentum from $p_T$ broadening, the final kinetic restriction on the radiated gluons that fall outside jet cone is given by
\begin{equation}
|\vec l_T + \Delta \vec l_T|/ zE > \sin R. \label{cut:collinear.bro}
\end{equation}

It was shown in \cite{Vitev:2008rz} that in the soft gluon emission limit the radiative
energy loss of a jet is given only by the energy of the gluon
transported outside of the jet cone of radius $R$. We will work within
the high-twist model \cite{Guo:2000nz, Wang:2001ifa,Wang:2009qb}, in which we approximate parton energy
loss by limiting the $z$ integration to $\frac{1}{2}$ as the case in Eq.~\ref{eq:rad.lp}.
Including all three modifications mentioned above, one can get the radiative jet energy loss from a modified version of Eq.~(\ref{eq:rad.lp}),
\begin{eqnarray}
\frac{\Delta E_{\rm jet}^{\rm rad}}{E}& \approx & \frac{2}{\pi} 
C_A \alpha_s  \int d\tau \int_0^{0.5} dz \int \frac{dl_T^2}{l_T^4} ~~ \nonumber\\
& & \int d^2 \Delta \vec l_T \frac{1}{2\pi \langle \Delta l_T^2 \rangle} e^{-\Delta \vec l_T^2/(2\langle \Delta l_T^2 \rangle)}\nonumber\\
& & (1-\theta(zE-\mu_D) \theta(zE\sin R - |\vec l_T + \Delta \vec l_T|)) \nonumber\\
& & \times \hat q_a z P_{ga} (z) \sin^2 \left( \frac{l_T^2 \tau}{4z(1-z)E} \right).
\end{eqnarray}
The theta functions here make sure that hard gluon radiations inside of the jet-cone do not contribute to the jet energy loss.

\begin{figure} [htb]
\includegraphics[width=0.45\textwidth]{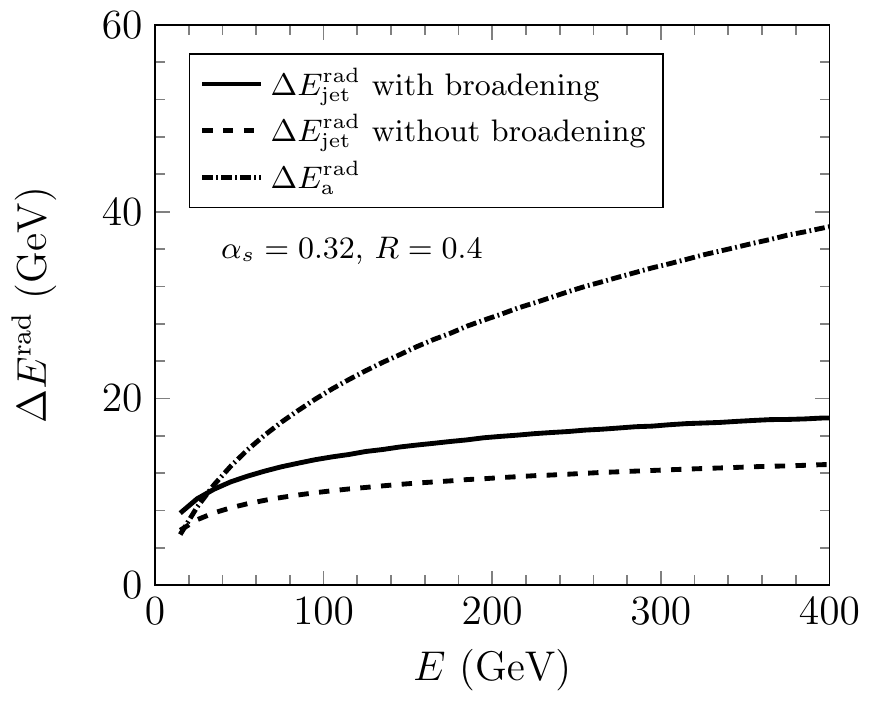}
\caption{
Radiative energy loss for a quark jet ($R=0.4$) with (solid) and without (dashed) $p_T$ broadening as a function of the initial jet energy as compared to the parton energy loss (dot-dashed). The initial jet is produced at $x=y=z=0$ and propagates along $\phi=0$ in $0-10\%$ central Pb+Pb collisions at $\sqrt{s_{\rm NN}} = 2.76$ TeV.
}
\label{fig:bro.unbro}
\end{figure}

Shown in Fig.~\ref{fig:bro.unbro} is the jet energy loss of a quark jet due to gluon bremsstrahlung with (solid) and without (dashed) transverse momentum broadening as a function of the initial jet transverse momentum. The initial configurations of jets and hydro profile are the same as that used in the calculations shown in Fig.~\ref{fig:radiative.and.collisional}. One can clearly see that the broadening effect is very significant which almost doubles the radiative jet energy loss. Overall, the jet radiative energy loss $\Delta E_{\rm jet}^{\rm rad}$ is much larger than the net jet energy loss from elastic scattering $E_{\rm jet}^{\rm el}$.
We also plot the energy loss of a leading parton (dot-dashed) for comparison. 
In the calculation of the radiative energy loss of the leading parton,
we need to introduce the Debye screening mass to regulate the collinear radiation
instead of the jet cone-size that is used in the calculation of the jet energy loss.

\begin{figure} [htb]
\includegraphics[width=0.45\textwidth]{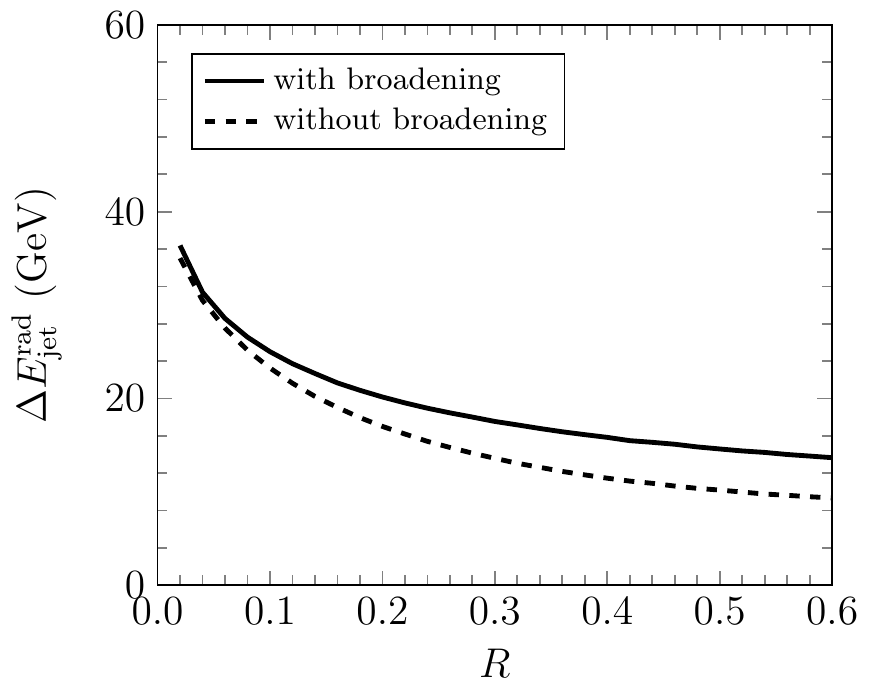}
\caption{
Radiative energy loss of a 200 GeV quark jet that is produced at the $x=0$, $y=0$, $\phi=0$ and propagates through medium in $0-10\%$ central Pb+Pb collisions at $\sqrt{s_{\rm NN}}=2.76$ TeV as a function of the jet cone-size.
}
\label{fig:cone-size.dependence}
\end{figure}

To study the cone-size dependence of the radiative jet energy loss, we show in Fig. \ref{fig:cone-size.dependence}
the radiative jet energy loss of a quark jet with initial energy  $E=200$ GeV with (solid) and without (dashed) transverse momentum broadening as a function of the jet cone-size $R$. The initial configurations of jets and hydro profile are the same as that for results shown in Fig. \ref{fig:radiative.and.collisional}. The jet energy loss in general decreases with the jet cone-size in both cases as more and more radiated gluons fall into the larger jet cone and do not contribute to the jet energy loss. With a finite jet cone-size, soft gluons whose energy is less than the value $zE< |\vec l_T+\Delta\vec l_T|/\sin R$ falls outside the jet cone and contribute to jet energy loss. This is why the jet energy loss does not vanish at very large cone-size. The cone-size dependence becomes weaker after transverse momentum broadening is taken into account.

\section{Jet suppression in high-energy heavy-ion collisions}

One of the direct consequences of jet energy loss is the suppression of the single inclusive jet cross section in
heavy-ion collisions relative to p+p collisions. We will calculate the suppression factor for the single inclusive jet
production in heavy-ion collisions within the NLO pQCD model in which the final jet energy is reduced from its initial value by the jet energy loss which is the sum of elastic and radiative jet energy loss.

In the central rapidity region, the initial transverse momentum of the produced jet comes mainly from 
the hard parton scattering and the intrinsic transverse momentum of the colliding partons can be ignored.
The single inclusive jet cross section in p+p collisions, $p+p \to {\rm jet} + X$, at large $p_T$, can be expressed
in a collinear factorized form at the leading order of pQCD as,
\begin{eqnarray}
\frac{d \sigma^{pp}}{dp_{T}dy} & = & 2 p_{T}\sum_{abcd}  \int dy_d  x_a f_{a/p} (x_a, \mu^2) x_b f_{b/p} (x_b, \mu^2) 
 \nonumber\\ 
 & &  \times \frac{d\hat\sigma_{ab\to cd}}{dt},
\label{eq:cs.pp}
\end{eqnarray}
where $y=y_c$ and $y_d$ are rapidities of the final partons in the $2\rightarrow 2$ processes, $x_a=x_T(e^y+e^{y_d})$, $x_b=x_T(e^{-y}+e^{-y_d})$ are the light-cone momentum fractions carried by the initial partons from the two colliding protons with $x_T=2p_T/\sqrt{s_{\rm NN}}$, $f_{a/p}(x,\mu^2)$ is the parton distribution function inside a proton and 
$d\hat\sigma_{ab\to cd}/dt$ is the parton level leading order cross section which depends on the Mandelstam variables 
$\hat s=x_ax_bs_{\rm NN}$, $\hat t=-p_T^2(1+e^{y_d-y})$ and $\hat u=-p_T^2(1+e^{y-y_d})$.

One assumes that the initial production rate of high $p_T$ jets in A+A collisions is the same as the superposition of nucleon-nucleon collisions,
except that one needs to consider jet energy loss and the nuclear modification of the initial parton distributions  \cite{Eskola:2009uj}.
When jet energy loss is considered, the cross section for single inclusive jet production in A+A collision at LO is given by,
\begin{eqnarray}
\frac{d \sigma^{AA}}{dp_{T}dy} & = &\sum_{abcd}  \int d^2r d^2b  t_A(r) t_A(|{\bf b}-{\bf r}|) \int \frac{d\phi}{\pi}  dy_d \nonumber\\
&& \times \left [ p_T x_a f_{a/A} (x_a, \mu^2) x_b f_{b/B} (x_b, \mu^2) \right .
\nonumber \\
&& 
\left .  \frac{d\hat\sigma_{ab\to cd}}{dt}\right ]_{p_T \rightarrow p_T+\Delta E^c_{\rm jet}},
\label{eq:cs.aa}
\end{eqnarray}
where $t_{A}(r)$ is the nuclear thickness function and $f_{a/A}(x,\mu^2)$ is the nuclear modified parton distribution function \cite{Eskola:2009uj},
$\bf r$ is the transverse coordinate of the binary nucleon-nucleon collision that produces the initial jet, $\bf b$ is the impact-parameter of the nucleus-nucleus collisions. The jet energy loss $\Delta E^c_{\rm jet} = \Delta E^{\rm el}_{\rm jet}+ \Delta E^{\rm rad}_{\rm jet}$ in the medium, which is described by a hydrodynamical model, depends on a path integral for given $\bf r$, $\bf b$ and the azimuthal angle $\phi$ of the jet. The range of integration over the impact parameter $\bf b$ is determined by the centrality of the nucleus-nucleus collisions according to the experimental measurement. Since the jet energy loss depends on the jet cone-size $R$, the jet cross section in A+A collisions will also depends on $R$ though the LO jet cross section in p+p collisions does not depend the jet cone-size.

The jet suppression factor is given by the ratio of the jet cross sections for A+A and p+p collisions normalized by the averaged number of binary nucleon-nucleon collisions,
\begin{equation}
R_{AA}=\frac{1}{\int d^2rd^2b  t_A(r) t_A(|{\bf b}-{\bf r}|)} \frac{d\sigma^{AA}}{d\sigma_{pp}}.
\end{equation}

\begin{figure}[htb]
\includegraphics[width=0.45\textwidth]{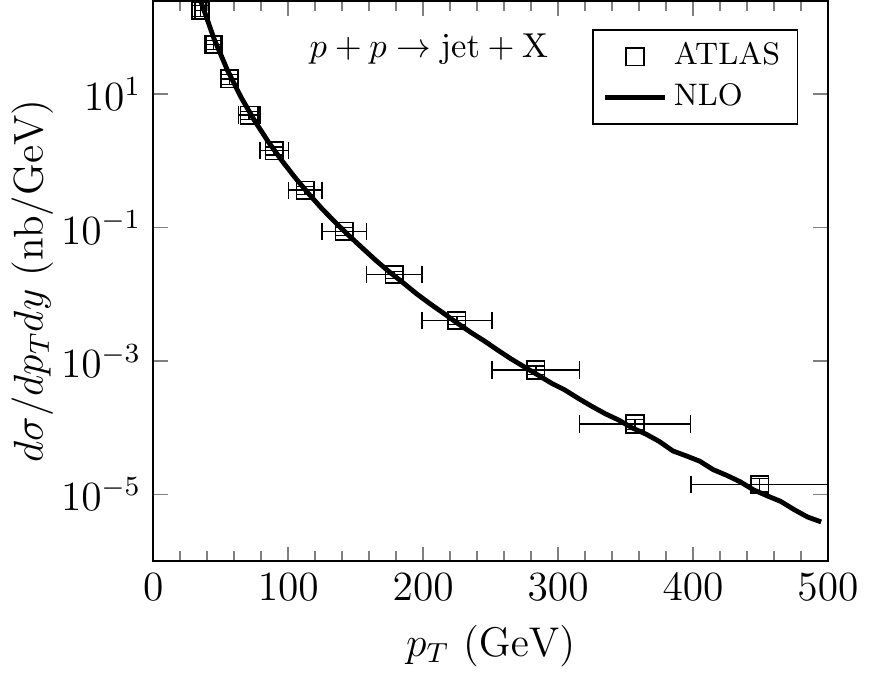}
\caption{Cross section for single inclusive jet production in p+p collisions at $\sqrt{s_{\rm NN}}=2.76$TeV from NLO pQCD calculation as compared to the experimental data by ATLAS experiment  \cite{Aad:2014bxa}.}
\label{fig:nlo.and.pp.exp}
\end{figure}

At NLO, one needs to include the $2\to 3$ parton scattering processes and virtual corrections to $2\to 2$ processes \cite{Harris:2001sx}.  The NLO pQCD model can describe well the experimental data on single inclusive jet cross section in p+p collisions at the LHC as measured by ATLAS \cite{Aad:2014bxa} and ALICE \cite{Adam:2015ewa}.  Shown in  Fig. \ref{fig:nlo.and.pp.exp} is the NLO pQCD result on single inclusive jet cross section in p+p collisions at $\sqrt{s_{\rm NN}}=2.76$ TeV as compared to the experimental data from ATLAS experiment \cite{Aad:2014bxa}.

We will also use NLO pQCD model to calculate the single inclusive jet cross section in heavy-ion collisions in which we include jet energy loss in the QGP medium similarly as in the LO case in Eq.~(\ref{eq:cs.aa}).  We therefore need to calculate the jet energy loss as the initial jet partons have to traverse the dense QGP medium before they hadronize and form the reconstructed jets as detected by experiments. 

\begin{figure}[htb]
\includegraphics[width=0.45\textwidth]{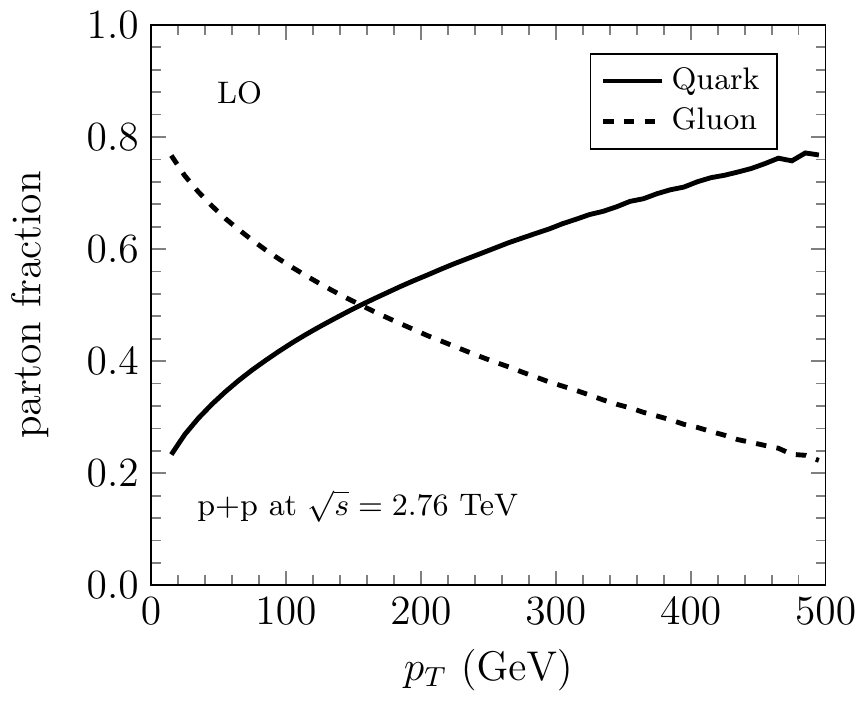}
\caption{Quark and gluon fractions in leading-order approximation in p+p collisions at $\sqrt{s}=2.76$ TeV.}
\label{fig:qog.LL}
\end{figure}

Jets produced in p+p or A+A collisions could be either a quark jet or a gluon jet. The energy losses for the quark jet and gluon jet in QGP medium are different. One therefore has to estimate the fractions of quark and gluon jets to calculate the jet cross section in heavy-ion collisions.  At leading order, the fractions can be calculated from pQCD parton model as shown in Fig. \ref{fig:qog.LL} where gluon jet dominates at low $p_T$ while quark jet dominates at high $p_T$. At the next-to-leading order, however,  the flavor of a jet becomes ambiguous since a jet can contain a leading parton and a gluon from final  state bremsstrahlung.  

In the NLO pQCD calculation, one has to combine two collinear partons into one single jet and sum up contributions from the virtual corrections to the single parton cross section in order to keep the final results infrared safe. In our calculation we approximate the flavor of a jet by that of the leading parton in a jet with a given cone-size which could contain two or more partons.

\begin{figure}[htb]
\includegraphics[width=0.45\textwidth]{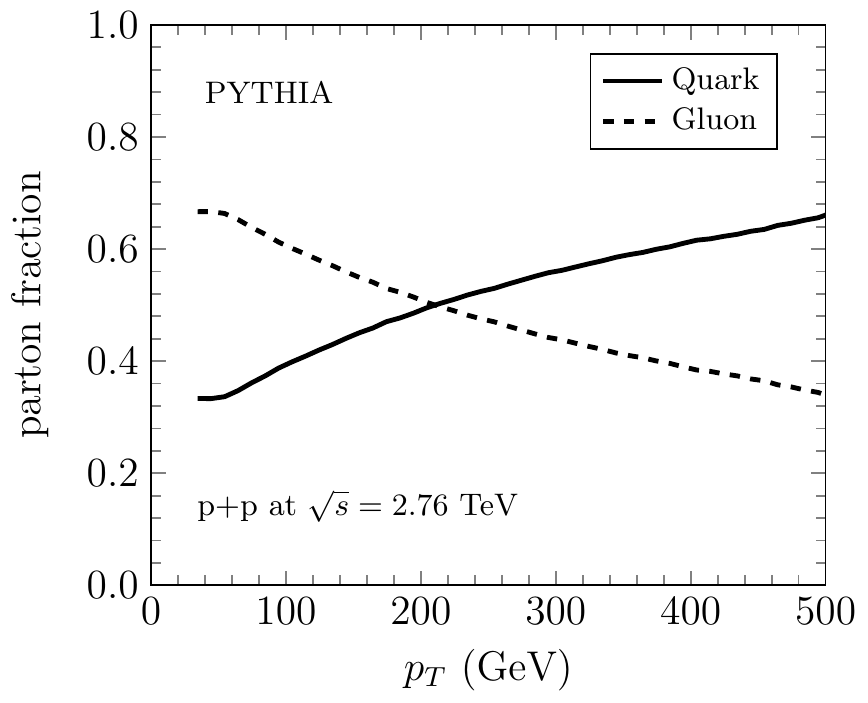}
\caption{Quark and gluon fractions with PYTHIA in p+p collisions at $\sqrt{s}=2.76$ TeV.}
\label{fig:qog.PYTHIA}
\end{figure}

Beyond the NLO in pQCD, one has also to consider parton shower in the final state evolution of a jet with high virtuality. In this case, 
parton splitting processes such as $q\to qg$, $g\to gg$ and $g\to q\bar q$ have to be considered.
Shown in in Fig. \ref{fig:qog.PYTHIA} are the quark and gluon jet fractions from  PYTHIA \cite{Sjostrand:2006za,Sjostrand:2007gs} simulations of p+p collisions at $\sqrt{s_{\rm NN}}=2.76$ TeV with cone-size $R=0.4$. Here we define the flavor the leading parton within the jet as the flavor of the jet. These fractions  are similar to the LO results in Fig. \ref{fig:qog.LL}. The difference between quark and gluon jet fractions from PYTHIA simulations is only slightly smaller than the LO results at both low and high $p_T$.  In our calculations of jet suppression factors in A+A collisions in the following, we will use both the LO and PYTHIA results on the quark and gluon jet fractions.

\begin{figure} [htb]
\includegraphics[width=0.45\textwidth]{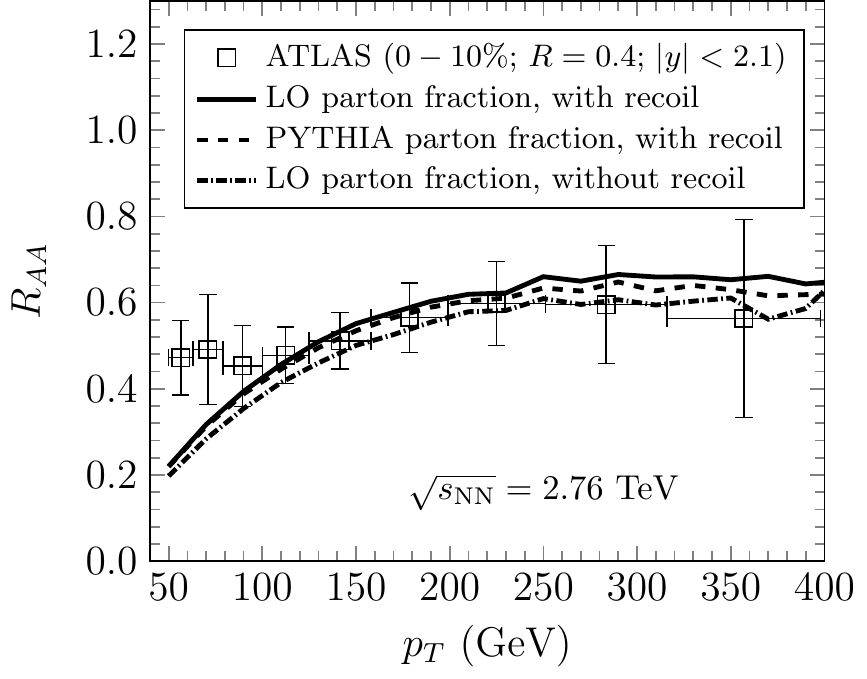}
\caption{
Jet suppression factor in central Pb+Pb collisions at $\sqrt{s_{NN}} = 2.76$TeV calculated with and without the recoil parton in the collisional energy loss, 
comparing with the ATLAS data \cite{Aad:2014bxa}. 
}
\label{fig:atlas.normal}
\end{figure}

Shown in Fig.~\ref{fig:atlas.normal} are our calculated suppression factors in central ($0-10\%$) Pb+Pb collisions at $\sqrt{s_{\rm NN}}=2.76$ TeV with both LO (solid) and PYTHIA (dashed) parton fractions and jet energy loss that include recoils partons within the jet cone and $p_T$ broadening of radiated gluons. The jet cone-size is $R=0.4$ and rapidity range of the jet is $|y|<2.1$. The difference in the suppression factors with two different parton fractions is very small.  Our results compare well with the  ATLAS data \cite{Aad:2014bxa} as shown in the figure, except at lower jet energy. We also show the suppression factor due to jet energy loss without the recoil partons as the dot-dashed line. Since the elastic part of the jet energy loss without recoil partons is larger than that with recoil partons, the corresponding jet suppression factor is also slightly smaller. Since the radiative energy loss is more dominant than the elastic jet energy loss, the effect of the recoil partons on the total jet energy loss is not very big. Their effect on the jet suppression factors are also small. The effect of parton broadening on the jet energy loss is quite big as we shown in the last section, its effect on jet suppression factor should also be very large. In principle, one should also consider the $p_T$ broadening for recoil partons. Since the elastic energy loss is small compared to the radiative jet energy loss, the effect of $p_T$ broadening of recoil partons should also be very small and can be neglected.

\begin{figure} [htb]
\includegraphics[width=0.45\textwidth]{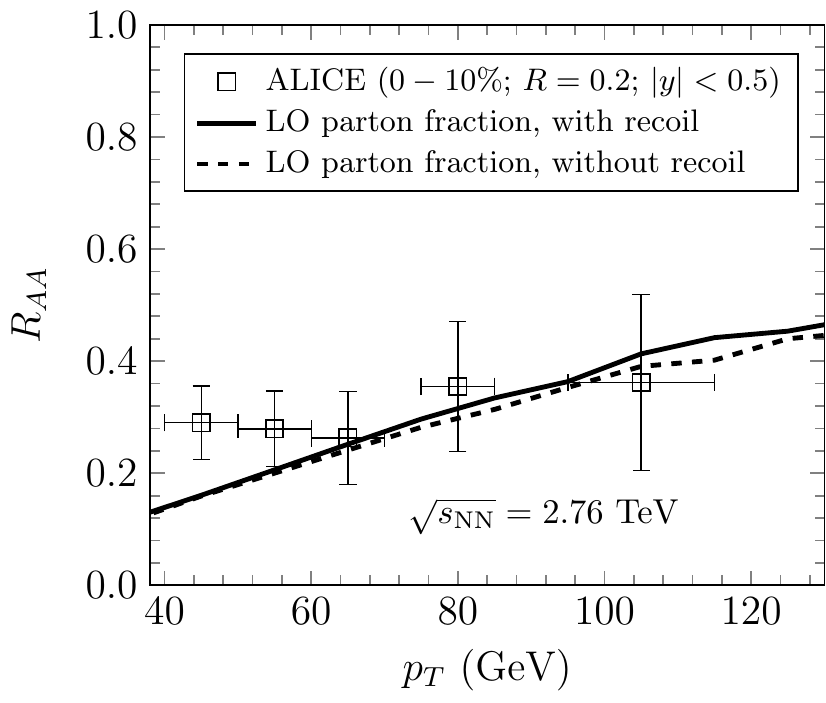}
\caption{
Jet suppression factor in central Pb+Pb collisions at $\sqrt{s_{NN}} = 2.76$TeV calculated with and without recoil parton in the collisional energy loss, 
comparing with the ALICE data \cite{Adam:2015ewa}.
}
\label{fig:alice.normal}
\end{figure}

We also calculate the jet suppression factor in central ($0-10\%$) Pb+Pb collisions at $\sqrt{s_{\rm NN}}$ with smaller jet cone-size $R=0.2$
and smaller rapidity range $|y|<0.5$ and compare to the ALICE data in Fig.~\ref{fig:alice.normal}. The recoil partons and $p_T$ broadening of radiative partons are both considered in the calculation of the jet energy loss. Because of the smaller jet cone-size, the jet energy loss should be bigger and therefore the jet suppression factor is smaller than that with a large jet cone-size $R=0.4$ as shown in Fig.~\ref{fig:atlas.normal}. Our results compare well again with the ALICE data. At lower $p_T$, the jet suppression factors from our  theoretical calculations are smaller than the experimental data for both jet cone-sizes. This might be caused by other effects that we have not considered in our study such as further transport of radiated gluons and recoil partons and their contribution to the jet energy within the jet cone.

\begin{figure} [htb]
\includegraphics[width=0.45\textwidth]{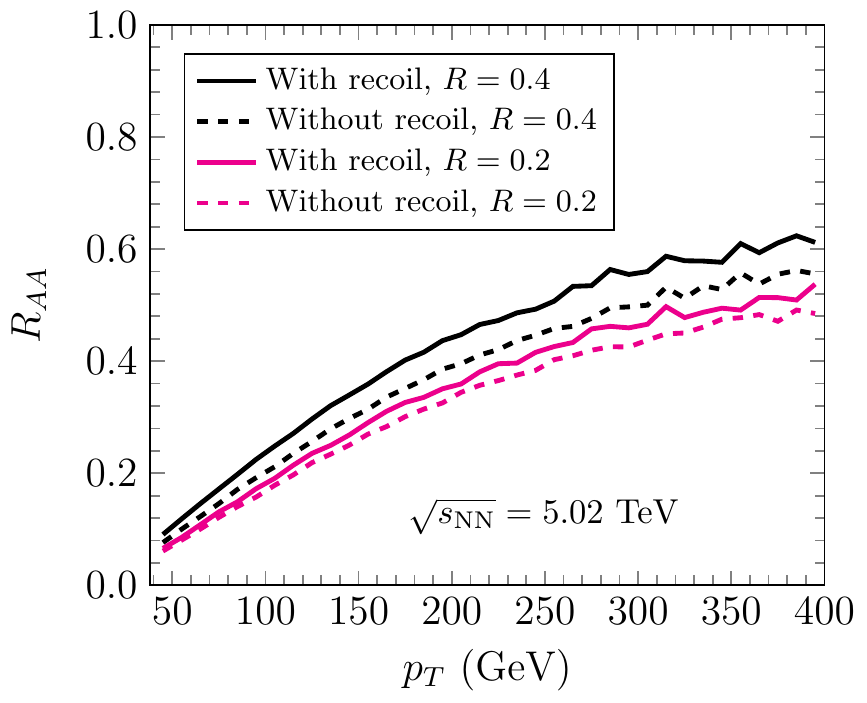}
\caption{
Jet suppression in central Pb+Pb collisions at 5.02 TeV LHC for
$R=0.4$ and $R=0.2$ in the rapidity region $|y|<2.1$.
}
\label{fig:5.02.Raa}
\end{figure}

Finally we show our prediction for the jet suppression factor $R_{AA}$ in central  ($0-10\%$) Pb+Pb collisions at $\sqrt{s_{\rm NN}}=5.02$ TeV with a jet cone-size $R=0.4$ or $R=0.2$ in rapidity region $|y|<2.1$. The hydro profile for Pb+Pb collisions at this energy we used in our calculation of jet energy loss is provided by L. G. Pang \cite{Pang:2012he,Pang:2014ipa} with initial temperature at the center of the collisions $T_0= 478.5$ MeV.  The jet suppression factors shown in Fig. \ref{fig:5.02.Raa} are calculated with (solid) and without (dashed) inclusion of recoil partons in the calculation of jet energy loss. 

\section{Summary}

In this paper, we studied the effect of medium recoil and $p_T$ broadening to the jet energy loss.
With a given jet cone-size, inclusion of recoil partons is found to reduce the elastic jet energy loss by a factor of 2.
Within the high-twist approach to parton energy loss,  radiated gluons induced by parton-medium interaction
are most likely to be collinear with the the leading  parton and therefore do not contribute much to the jet energy loss with a given cone-size.
We find that inclusion of $p_T$ broadening of radiated gluons 
{
using a Gaussian broadening model
} 
can transport radiated gluons to outside jet-cone and can increase significantly the jet radiative energy loss. 
Our calculated nuclear suppression factor $R_{AA}$ caused by jet energy loss including both recoil partons and $p_T$ broadening of radiated gluons can describe experimental data at LHC ($\sqrt{s_{\rm NN}}=2.76$ TeV) well. We also calculated the jet suppression factor in Pb+Pb collisions at $\sqrt{s_{\rm NN}}=5.02$ TeV.

\section*{Acknowledgements}
We thank Y. He for providing us with quark and gluon fractions from PYTHIA simulations and L. G. Pang for providing the hydro profiles in Pb+Pb collisions at LHC.
We would like to thank N.B. Chang, G.Y. Qin and B.W. Xiao for helpful discussions.
This work is supported by the Natural Science Foundation of China (NSFC) under Grant No. 11435004 and No. 11521064, by the Chinese Ministry of Science and Technology under Grant No. 2014DFG02050, by the Major State Basic Research Development Program in China under contract No. 2014CB845400 and No. 2014CB845404 and by the Director, Office of Energy Research, Office of High Energy and Nuclear Physics, Division of Nuclear Physics, of the U.S. Department of Energy under Contract No. DE-AC02-05CH11231,

\end{document}